\documentclass[twocolumn, showpacs,preprintnumbers,amsmath,amssymb]{revtex4}
\usepackage[dvips]{graphicx}
\usepackage{graphics, lscape, afterpage, longtable, bm,
}
\begin{document}
\title{Thermodynamics of Ion Solvation and Differential Adsorption at Liquid-Liquid Interfaces and Membranes }
\author{William Kung}\email{w-kung@northwestern.edu}
\affiliation{Department of Materials Science and Engineering, Northwestern University, Evanston, Illinois 60208-3108}
\author{Francisco J. Solis}\email{francisco.solis@asu.edu}
\affiliation{Department of Integrated Natural Sciences, Arizona State University, Glendale, Arizona 85306}
\author{Monica Olvera de la Cruz}\email{m-olvera@northwestern.edu}
\affiliation{Department of Materials Science and Engineering,
Department of Chemistry, Department of Chemical and Biological
Engineering, Northwestern University, Evanston, Illinois 60208-3108}

\begin{abstract}
We construct a mean-field formulation of the thermodynamics of ion
solvation in immiscible polar binary mixtures.  Assuming an
equilibrium planar interface separating two semi-infinite regions of
different constant dielectric medium, we study the electrostatic
phenomenon of differential adsorption of ions at the interface.
Using general thermodynamic considerations, we construct the
mean-field $\Omega$-potential and demonstrate the spontaneous formation of an
electric double-layer around the interface necessarily follow.  In
our framework, we can also relate both the bulk ion densities in the
two phases and the distribution potential across the interface to
the fundamental Born free energy of ion polarization.  We further
illustrate this selective ion adsorption phenomenon in respective
examples of fully permeable membranes that are neutral, negative, or
positive in charge polarity.
\end{abstract}

\date{\today}

\maketitle

\def\coth{\hbox{\rm{coth}}}
\def\csch{\hbox{\rm{csch}}}
\def\sech{\hbox{\rm{sech}}}

\twocolumngrid
\section{Introduction}

Electro-chemical and physical  processes at liquid-liquid interfaces
and across membranes are broadly important in many systems. Examples
range from cellular and physiological systems~\cite{Po,New} to
everyday applications, such as portable batteries, as well as
scenarios for the origin of life~\cite{Onsager} and also
biotechnology~\cite{Sikorav}. Biologically, ion transport is
essential for transmembranous and transcellular electric potential,
fluid transport, and maintaining cellular volume~\cite{Al`	bertsText}; the failure
in the regulation of the above would lead to such conditions as
septicemia induced pulmonary edema, electrolyte abnormalities in
pyelonephritis of early infancy, hypovolemia and
hyponatremia~\cite{Eisenhut2006}.  In the chemistry community, there
has been a long interest in the rich electrochemistry associated
with interfacial charge transfer that is important in, for example,
hydometallurgy and two-phase electrolysis~\cite{Samec1988}.

The interaction of ionic species, particularly proteins and nucleic
acids, with interfaces and membranes has become exceedingly important in our understanding of 
biological systems~\cite{Po} and in processes such as genetic
transformation, since the latter specifically involves DNA crossing cellular membranes.
Moreover, important biomolecular processes take place more efficiently at
heterogeneous media than in homogeneous systems, as demonstrated in the example of DNA renaturation at water-phenol interfaces~\cite{Sikorav}.
Therefore, the study of ionic profiles at interfaces also leads to a
better understanding of these various biomolecular processes. In general, {\it{in situ}}
experimental techniques are required to evaluate interface
phenomena. For example, x-ray-standing waves measurements provide the profile of
ions~\cite{Bedzyk} and also more recently of nucleic acids~\cite{Hao}, both along
hard surfaces in aqueous solutions~\cite{Netz2003}. On the other hand, recent experimental techniques
have detected ion density profiles at liquid-liquid interfaces with the use of x-ray reflectivity measurements~\cite{Schlossman2006a,
Schlossman2006b}. These experiments provide direct knowledge of
electrostatic interactions among ionic species at interfaces and can be easily extended to analyze the adsorption of proteins and nucleic acids
at liquid-liquid interfaces.

In their experimental  study~\cite{Schlossman2006a,
Schlossman2006b}, Luo {\it{et al.}} demonstrated that the
predictions from Gouy-Chapman theory did not match well with x-ray
reflectivity measurements, presumably due to its neglect of
molecular-scale structure in the liquid solution.  To better fit
their experimental data and the free-energy profile of the ions, the authors instead formally introduce an
{\it{ad hoc}} term in the electrostatic energy, which presumably includes effects
from short-range correlations and which can be theoretically approximated
by the potential of mean
force, experienced by single ions near the interface, obtainable from molecular dynamics (MD) simulations.  Notwithstanding the absence of a fundamental  theoretical framework for the aforementioned
system of a planar liquid-liquid interface, progress has been made with phenomenological approaches, including the example of a Ginzburg-Landau theory of ions solvation and their distribution around an interface~\cite{Onuki}. 

In this work, we will present a top-down thermodynamic formulation for the system of
dissolved ions in immiscible polar-binary mixtures.  In our formulation, we can
naturally incorporate details such as the interaction between ions
and background solvents, at the mean-field level, via the Born
approximation~\cite{LyndenBell2001, LyndenBell1997}.  We will then
relate the various hitherto phenomenological parameters, namely, the
bulk ion densities in each phase as well as  the distribution
potential across the interface, to the fundamental parameters of our
system that include the dielectric constant of each phase and the
size and charge of dissolved ions.  In addition, the full nonlinear
Poisson-Boltzmann equation follows from variational principle and
can be analytically solved for the electrostatic potential and for
the ion density profiles without further approximations.
Electro-neutrality, in our formulation, simply manifests from
consistency requirement with thermodynamics.

As an important  consequence of thermodynamics and
electrostatics, we will demonstrate the physics of
{\it{differential adsorption}} of ions near an interface, which
is a general phenomenon occurring between two phases of
different dielectric medium wherein ions are selectively driven
into confinement near, or expulsion from, the interface based on
their charge polarity.  In particular, we will consider examples
involving a neutral membrane, a negatively charged membrane, as
well as a positively charged one partitioning two chemically
different environments.  Given the long-standing interest in this
topic across diverse disciplines and the many still unresolved
issues, it is inevitable that there exists many prior works on
the subject with very different emphases and approaches.  For
example, chemists have traditionally focussed on detailing the
microscopic molecular interactions between ions, solvents, and
the different types of membranes~\cite{Samec1988, Karpfen1953,
Donnan1911}.  For the same reason, there exists considerable
varied usage of terminology~\cite{Markin1989} regarding similar concepts.  For clarity, we will henceforth explicitly
consider in this work only permeable membranes.  In addition, to
establish notations and conventions, we will refer to the
potential gradient across the interface, due to the different
existing bulk ions densities in the two phases, as the
distribution potential $\Phi_D$.  The cases of neutral and
charged membranes (regardless of polarity) would correspond to
the terminology of nonpolarizable and polarizable liquid-liquid
interfaces, respectively, as discussed in chemistry literature.  While
partial results regarding these systems have been derived in
various contexts and guises,  our present work represents a
unifying framework in a complete, mean-field thermodynamic
formulation for the system of dissolved ions in immiscible polar binary
mixtures, within which we can derive rigorous relations between the
different phenomenological parameters of the system that would
otherwise be unrelated when treated in the context of
electrostatics alone.

In what follows,  we will present, in section II, our
thermodynamic formalism in terms of a mean-field
$\Omega$-potential and derive, via variational principle, the
relevant electrostatics and thermodynamic constraints in our
system.  We will then proceed in section III to consider full
solutions to the nonlinear Poisson-Boltzmann equation in this new
context and demonstrate the general mechanism of differential
adsorption of ions across a liquid-liquid interface.  Section IV
concludes our paper.

\begin{figure}[t]
\vspace{2mm}
\hspace{0mm}
\includegraphics{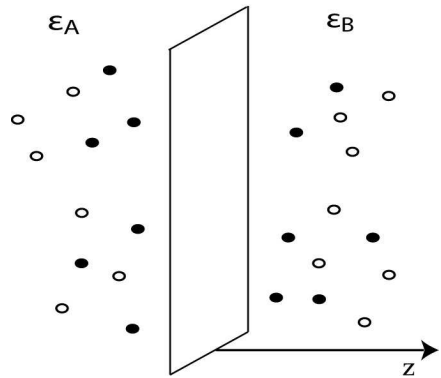}
\vspace{2mm}
\caption{A single charged plate in salt solution.  The system is symmetric with respect to $z=0$.  Without loss of generality, the charged plate is assumed to have a positive surface-charge density $\sigma$, with counterions (solid circles) and coions (open circles) present on both sides of the plate.}
\label{oneplate}
\end{figure}

\section{Thermodynamics of Ion Solvation}

For the system of  dissolved salt ions in polar binary mixtures,
there are several parameters that have been, as previously mentioned,
traditionally considered as independent {\it{inputs}} to the
formulation of such systems.  These input parameters include, for
each phase $\alpha$, the dielectric constants
$\varepsilon_{\alpha}$, the bulk ion densities
$\rho^{\alpha}_{0,\pm}$, and the distribution potential across the
interface $\Phi_D$.  To further characterize these parameters, we
now turn to thermodynamic considerations.  We will henceforth denote
the two polar phases by $\alpha = A, B$ with $A$ labeling the
more polar phase.  In principle, the condition of equilibrium
implies equal chemical potential across the interface,
$\mu^A_{\pm}=\mu^B_{\pm}$, which should in turn determine the bulk
densities of the salt ions existing in each of the two phases.  The
thermodynamics with this equilibrium condition are formulated by
constructing the $\Omega$-potential (equals to $-PV$ where $P$ is the
pressure and $V$ is the volume) for each phase $\alpha$ as follows:
\begin{eqnarray}
\Omega^{\alpha}&=&\Omega^{\alpha}_{el}+\Omega^{\alpha}_{th}+\Omega^{\alpha}_{sol},
\label{omega}
\end{eqnarray}
where we have divided  the potential into a contribution due to
electrostatics $\Omega_{el}$, to thermodynamics $\Omega_{th}$,
and to solvation interactions $\Omega_{sol}$.  In what follows, all energy quantities will be measured in units of $k_BT$, all length scales in units of the Bjerrum length $\ell_B={\bf{\mathrm{e}}}^2/k_BT$, and all ionic charges quoted as multiples of the fundamental electron charge ${\bf{\mathrm{e}}}$.  The electrostatic
part delineates the free-energy dependence on the dielectric
constant $\varepsilon^{\alpha}$ and imposes Gauss' law via a
Lagrange-multiplier constraint using the electrostatic potential
$\Phi$:
\begin{eqnarray}
\Omega^{\alpha}_{el}&=&\int\,\frac{\epsilon^{\alpha}\vert\bm{E}^{\alpha}(\bm{x})\vert^2}{8\pi}\nonumber\\
&&-\int\,\Phi^{\alpha}\left(\bm{\nabla}\cdot\frac{\epsilon^{\alpha}\bm{E}^{\alpha}(\bm{x})}{4\pi}-\rho^{\alpha}_{+}(\bm{x})-\rho^{\alpha}_{-}(\bm{x})\right).\nonumber\\
\end{eqnarray}
On the other hand, the  thermodynamic contribution consists of
the entropy associated with the ions, treated as point particles,
in solution as well as their corresponding chemical potentials
$\mu^{\alpha}_{\pm}$,
\begin{eqnarray}
\Omega^{\alpha}_{th}&=&\rho^{\alpha}_{+}(\bm{x})\log\left[\frac{\rho^{\alpha}_{+}(\bm{x})}{e}\right]+\rho^{\alpha}_{-}(\bm{x})\log\left[\frac{\rho^{\alpha}_{-}(\bm{x})}{e}\right]\nonumber\\
&&-\mu^{\alpha}_+\rho^{\alpha}_{+}(\bm{x})-\mu^{\alpha}_-\rho^{\alpha}_{-}(\bm{x}),
\end{eqnarray}
where the chemical  potentials will be specified by the
equilibrium constraint that they should be equal across the
interface.  Lastly, the solvation interaction between the ions
and the respective polar solvent in each phase is modeled by the
mean-field Born approximation of polarziation;
\begin{eqnarray}
\Omega^{\alpha}_{sol}&=&-g^{\alpha}_+\rho^{\alpha}_{+}(\bm{x})-g^{\alpha}_-\rho^{\alpha}_{-}(\bm{x}),
\end{eqnarray}
where the Born  polarization energy $g^{\alpha}_{\pm}$ is given
by
\begin{eqnarray}
g^{\alpha}_{\pm}&=&\frac{Z^2_{\pm}{\bf{{\mathrm{e}}}}^2}{8\pi\varepsilon^{\alpha}R_{\pm}},
\end{eqnarray}
for ions with valency $Z_{\pm}$ and radius $R_{\pm}$.  As
expected, we note that an increase in dielectric constant
$\varepsilon^{\alpha}$ or in the ionic radius $R_{\pm}$ for a
given valence $Z_{\pm}$ facilitates solvation of ions in polar
mixtures and leads to an overall lower Born energy.  Consequently, Born's model predicts a larger
solubility for anions (which are generally larger in size due to
reduction in redox chemistry) than for the corresponding cations.
As we will see in subsequent sections, both charge properties and
the size factor play an important role in the physics of ion
solvation near a boundary  interface.

To examine the consequence  of our thermodynamic formulation of ion
solvation in immiscible polar binary mixtures, we first perform dimensional
analysis on the electrostatic term.  We note that the
$\Omega$-potential is an extensive quantity by construction:
$\Omega^{\alpha}_{el}\propto V$, where $V$ is the volume of the
overall system.  Defining the total charge density
$\rho_{c}^{\alpha}(\bm{x})=\vert
Z_+\vert\rho^{\alpha}_{+}(\bm{x})-\vert
Z_-\vert\rho^{\alpha}_{-}(\bm{x})$, we assume that it has a
nonvanishing zero mode and observe that its contribution to the
energy density is
$\left(\rho_{\bm{0}c}^{\,\alpha}\right)^2\int\,d\bm{x}\int\,d\bm{y}\frac{1}{\vert\bm{x}-\bm{y}\vert}$.
Based on the dimensionality of the various factors within the
integral, it is straightforward to see that
$\Omega^{\alpha}_{el}\vert_{\bm{k}=\bm{0}}\propto V^{5/3}$, for
$\rho_{\bm{0}c}\neq 0$, which is inconsistent with the thermodynamic
construction of the extensive $\Omega$-potential. Therefore, the
requirement of consistency with equilibrium thermodynamics
automatically imposes charge neutrality in the overall system;
namely, $\rho^{\alpha}_{\bm{0}}=0$ (we note that this elelctroneutrality
condition in linearized approaches appears as a restriction to avoid
the divergence in the potential~\cite{Ermoshkin}).

From Eq. (\ref{omega}), variation with respect to the electric field $\bm{E}^{\alpha}$, the Lagrange multiplier (electrostatic potential) $\Phi^{\alpha}$, and the number density $\rho^{\alpha}_{\pm}$ readily yields the relations
\begin{eqnarray}
\label{p1}
\delta{\bm{E}}^{\alpha}\,&:& \,\,\,\,\, {\bm{E}}^{\alpha}(\bm{x})=-\bm{\nabla}\Phi^{\alpha}(\bm{x}),\\
\label{p2}
\delta\Phi\,&:& \,\,\,\,\, \bm{\nabla}\cdot\left[\frac{\varepsilon^{\alpha}{\bm{E}}^{\alpha}(\bm{x})}{4\pi}\right]=\rho^{\alpha}_{0+}(\bm{x})+\rho^{\alpha}_{0-}(\bm{x}),\\
\delta\rho_i^{\alpha}\,&:& \,\,\,\,\, \log\rho_{\pm}^{\alpha}(\bm{x})\pm\Phi^{\alpha}(\bm{x})-g^{\alpha}_{\pm}=\mu^{\alpha}_{\pm}.
\label{p3}
\end{eqnarray}
We note that Eqs. (\ref{p1})  and (\ref{p2}) simply reproduce,
respectively, the expressibility of the electric field $\bm{E}$
in terms of a scalar potential $\Phi^{\alpha}$ and the Gauss' law in
electrostatics.  The last relation in Eq. (\ref{p3}) provides an
explicit expression of the chemical potential for each phase in
terms of its respective ${\it{input}}$ parameters such as the dielectric constants and ion sizes.  It is now
obvious that the thermodynamic condition of equal chemical
potentials across the different phases $\alpha$ would provide
further constraints on the respective bulk ionic densities.  To
make connection with the distribution potential $\Phi_D$, we also
need to introduce explicitly now the further assumption of
Boltzmann distribution for the ion densities.  In particular, we
have, for the more polar phase $A$,
\begin{eqnarray}
\rho^{A}_{\pm}(\bm{x})&=&\rho^{A}_0\,e^{\mp\Phi^{A}(\bm{x})}e^{\pm\Phi_D},
\label{Boltzmann}
\end{eqnarray}
while a similar expression  (modulus the exponential factor of
the distribution potential $\Phi_D$) applies to the less polar
phase $B$.
Taking the logarithm of Eq.  (\ref{Boltzmann}) readily yields
\begin{eqnarray}
\log\rho^{A}_{\pm}(\bm{x})\pm\Phi^{A}(\bm{x})&=&\log\rho^{A}_{0}\pm\Phi_D
\label{91}
\end{eqnarray}
Again, a similar expression  to Eq. (\ref{91}), modulus the term
containing the distribution potential, applies to the less polar
phase $B$.  Now upon substituting Eq. (\ref{91}) and the
corresponding expression for phase $B$ into Eq. (\ref{p3}), as
well as imposing equal chemical potentials across boundary, we
obtain the following relations
\begin{eqnarray}
\log\rho^A_0+\Phi_D-g^A_+&=&\log\rho^B_{0}-g^B_+,\\
\log\rho^A_0-\Phi_D-g^A_-&=&\log\rho^B_{0}-g^B_-.
\end{eqnarray}
We can now take the sum and  difference of the two equations to
obtain the relations
\begin{eqnarray}
\log\rho_{0}^A+\bar{g}_A&=&\log\rho_{0}^B+\bar{g}_B,\\
2\Phi_D+\Delta g_A&=&\Delta g_B,
\end{eqnarray}
where we have defined the  average Born polarization energy over
ions of both charge polarity, $\bar{g}_{\alpha}=g_{\alpha
+}+g_{\alpha -}$, and the corresponding differential Born energy,
$\Delta g_{\alpha}=g_{\alpha +}-g_{\alpha -}$ between the cations
and anions.  In terms of these two quantities, we can relate the
bulk ion densities and the distribution potential to the
dielectric constants $\varepsilon_{\alpha}$ in each phase via
\begin{eqnarray}
\label{bulk}
\frac{\rho_0^A}{\rho_0^B}&=&e^{\bar{g}_B-\bar{g}_A}=e^{-\Delta_{\alpha}g}\\
\Phi_D&=&-\frac12\Delta_{\alpha}\Delta g
\label{lambda}
\end{eqnarray}
where $\Delta_{\alpha}\Delta g=\Delta g_A-\Delta g_B$.

In summary, we have thus  shown in Eq. (\ref{bulk}) that the bulk
densities $\rho^{\alpha}_{0,\pm}$ are related to the difference
of the average Born solvation energy between the two phases,
while in Eq. (\ref{lambda}) the distribution potential $\Phi_D$
is related to the difference in the differential Born energy
between ion types in the two phases.  We note that even though
the above formulation has been done for a neutral liquid-liquid
interface wherein the charge distribution of the system resides
completely amongst the dissolved ions in the bulk, it is
straightforward to generalize to the case where a surface
charge exists on the interface itself.

To determine the equilibrium  ion density profiles
$\rho^{\alpha}_{\pm}(\bm{x})$ as well as the electrostatic
potential $\Phi(\bm{x})$, it remains for us to solve the Poisson
equation [Eq. (\ref{p2})] after applying the integrability
condition [Eq. (\ref{p1})] as well as the thermodynamic ansatz of the
Boltzmann distribution for the ion density [Eq.
(\ref{Boltzmann})].  We will illustrate with two classes of
examples in the subsequent section involving non-polarizable
(neutral) but permeable membranes as well as polarizable
(charged) membranes.
\begin{figure}[t]
\begin{center}
\includegraphics{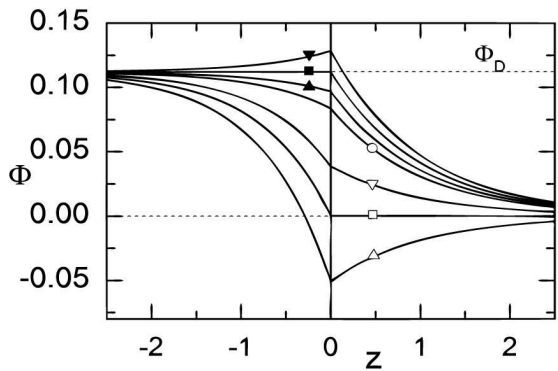}
\caption{Comparison of the electrostatic potential across a positively and negatively charged membrane in nitrobenzene/water solution at 68F.  The surface charge densities are quoted in units of $\epsilon_A\,\kappa_A$ and $\Phi'_D\approx 0.112$. The potential is plotted for $\sigma = -1.5 (\triangle)$, $\sigma = -0.93 (\Box)$, $\sigma= -0.5 (\triangledown)$, $\sigma=0 (\circ)$, $\sigma = 0.15 (\blacktriangle)$, $\sigma = 0.32 (\blacksquare)$, $\sigma= 0.5 (\blacktriangledown)$.  }
\label{potential1}
\end{center}
\end{figure}

\section{Electrostatics of Differential Adsorption}

As mentioned in the Introduction, it is now feasible to make
precise experimental detection of the ion distributions near
liquid-liquid interfaces~\cite{Schlossman2006a, Schlossman2006b}.
It has been found that while the mean-field Poisson-Boltzmann
framework provides a generally good description for the ion
profiles that matches well with experimental results obtained by
x-ray structural measurements, the linearized Guoy-Chapman
approximation is insufficient for most general
cases~\cite{Schlossman2006b}.  For completeness, we will now
illustrate the solution to the Poisson-Boltzmann equation for the case 
when the membrane across the two dielectric media is charged
(polarizable), including the particular interesting case of
the neutral (non-polarizable) interface. Furthermore, our
liquid-liquid interface is treated as an infinite two-dimensional
plane so that we do not need to consider surface terms.  We thus
essentially reduce our system to a one-dimensional problem.
Denoting the lateral distance from the membrane by $z$, we will
now formulate the full non-linear Poisson-Boltzmann equation with
the proper boundary conditions and obtain the corresponding
electrostatic potential $\Phi^{\alpha}(z)$ and ion density
profiles $\rho^{\alpha}_{\pm}(z)$ in each phase $\alpha$.
\begin{figure*}[t]
\begin{center}
\includegraphics{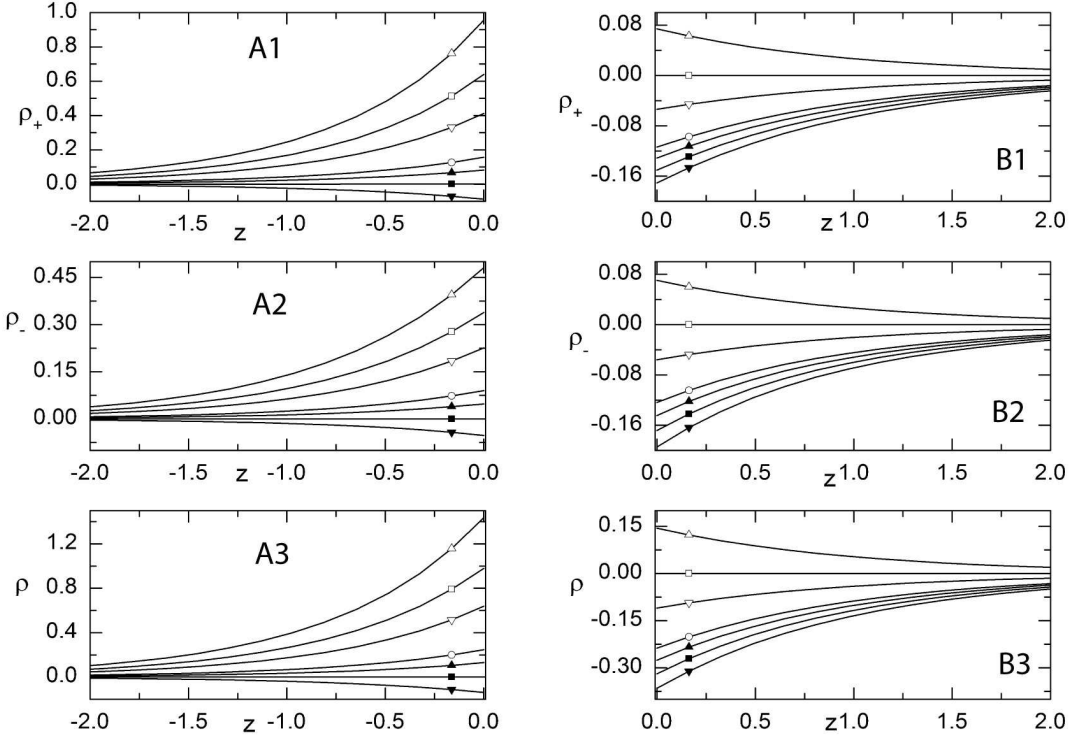}
\vspace{2mm}
\caption{Comparison of the ion distributions of ions in the two polar phases, $A\, (z<0)$ and $B\, (z>0)$ characterized by $\varepsilon_A$ and $\varepsilon_B$, respectively, when the surface density of the interface is $\sigma = -1.5 (\triangle)$, $\sigma = -0.93 (\Box)$, $\sigma= -0.5 (\triangledown)$, $\sigma=0 (\circ)$, $\sigma = 0.15 (\blacktriangle)$, $\sigma = 0.32 (\blacksquare)$, $\sigma= 0.5 (\blacktriangledown)$.   (1) Cations.  (2) Anions.  (3) Total charge density $\rho = \rho_++\rho_-$.  The curves are presented in separate plots because of difference in concentrations.  For this particular plot, we use a range of surface charge densities $\sigma$ in units of $\epsilon_A\,\kappa_A$ and $\Phi'_d=0.112$ for the nitrobenzene/water solution at 68F. }
\label{Profiles_neg}
\end{center}
\end{figure*}
%


To begin, we analytically formulate our system in which the two phases
$A$ and $B$ are in equilibrium, where each phase is characterized by its respective constant
dielectric permittivity $\varepsilon_A$ and $\varepsilon_B$, as
follows:
\begin{eqnarray}
\varepsilon_A\bm{\nabla\cdot E^A}&=&\rho^A_c,\,\,\,\,\,\,\,\,\,\,z>0\;,\\
\varepsilon_B\bm{\nabla\cdot E^B}&=&\rho^B_c,\,\,\,\,\,\,\,\,\,\,z<0\;,
\end{eqnarray}
\\
subject to the following boundary conditions at $z=0$:
\\
\begin{eqnarray}
\left.\varepsilon_BE^B_z\right\vert_{z\rightarrow 0^-}\-\left.\varepsilon_AE^A_z\right\vert_{z\rightarrow 0^+}&=&4 \pi \sigma,\\
E^A_x\vert_{z\rightarrow 0^+}&=&E^B_x\vert_{z\rightarrow 0^-}\;,\\
E^A_y\vert_{z\rightarrow 0^+}&=&E^B_y\vert_{z\rightarrow 0^-}\;.
\end{eqnarray}
\\
Making use of Eq. (\ref{p1}) and substituting the Boltzmann
distribution for ion density profile [Eq. (\ref{Boltzmann})] , we
now take advantage of the one-dimensional nature of our system
and collapse the above formulation into a single equation,
written in terms of the electrostatic potential $\Phi^{\alpha}(z)$, to
obtain the Poisson Boltzmann-equation representation of the Gauss'
law:
\begin{eqnarray}
\label{kAinhomo}
\frac{d^2\Phi^A}{dz^2}&=&\left(\kappa^A\right)^2\sinh\left(\Phi^A-\Phi_D\right)\;,\,\,\,\,\,\,\,\,\,\,z<0,\\
\frac{d^2\Phi^B}{dz^2}&=&\left(\kappa^B\right)^2\sinh\Phi^B\;,\,\,\,\,\,\,\,\,\,\,\,\,\,\,\,\,\,\,\,\,\,\,\,\,\,\,\,\,\,\,\,z>0,
\label{kBinhomo}
\end{eqnarray}
where, respectively, $\kappa^{\alpha}=\sqrt{4\pi
e^2\rho^{\alpha}_0/\epsilon^{\alpha} kT}$ is the inverse Debye
screening length, and
$\rho_0^{\alpha}=Z^2_+\rho^{\alpha}_{0,+}+Z^2_-\rho^{\alpha}_{0.-}$
is the combined bulk densities of the ions in the two phases,
$\alpha=A, B$.  The inverse Debye screening length
has been so defined as to scale the electrostatic potential
$\Phi^{\alpha}$ dimensionless.  The corresponding boundary
conditions for the potential are as follows:
\begin{eqnarray}
\label{0bc}
\left.\Phi^A\right\vert_{z\rightarrow 0^-}&=&\left.\Phi^B\right\vert_{z\rightarrow 0^+},\\
\label{0bc1}
\left.\varepsilon^A\frac{d\Phi^A}{dz}\right\vert_{z\rightarrow 0^-}-\varepsilon^B\frac{d\Phi^B}{dz}\vert_{z\rightarrow 0^+}&=&4 \pi \sigma,\\
\label{inftybc1}
\left.\frac{d\Phi^A}{dz}\right\vert_{z\rightarrow -\infty}&=&0,\\
\left.\frac{d\Phi^B}{dz}\right\vert_{z\rightarrow +\infty}&=&0.
\label{inftybc}
\end{eqnarray}
The general solution to the full nonlinear Poisson-Boltzmann equation can be readily found~\cite{Overbeek}.  At present, we simply furnish the particular solution for our system of nonpolarizable liquid-liquid interface specified by the above boundary conditions:
\begin{eqnarray}
\label{28}
\Phi^A(z)&=&\Phi_D-4\tanh^{-1}\left(C_A\, e^{\kappa_A z}\right)\;,\,\,\,\,\,z<0,\\
\Phi^B(z)&=&4\tanh^{-1}\left(C_B\, e^{-\kappa_B
z}\right)\;,\,\,\,\,\,\,\,\,\,\,\,\,\,\,\,z>0.
\label{29}
\end{eqnarray}
\\
As shown, Eqs. (\ref{28}) and (\ref{29}) are expressed in terms of the two integration
constants, $C_A$ and $C_B$, and as such, physically meaningful solutions would correspond to the range  of $|C_\alpha|\leq 1$.  Alternatively, we can rewrite $C_{\alpha}=\exp\left(-\kappa_{\alpha}z_{\alpha}\right)$, and the solutions now take the form of $\Phi^A(z)=\Phi_D-4\,{\rm{sgn}}(C_A)\,\tanh^{-1} e^{\kappa_A
(z-z_A)}$, and $\Phi^B(z)=4\,{\rm{sgn}}(C_B)\,\tanh^{-1} e^{-\kappa_B
(z-z_B)}$, where ${\rm{sgn}}[\,\cdot\,]$ is the sign function.  The
choice of signs in the above solutions is such that the integration constants are
positive when we have a neutral interface, $\sigma=0$.

In terms of the integration constants $C_A$ and $C_B$, the continuity equation in Eq. (\ref{0bc}) reads:
\begin{eqnarray}
\frac{(1+C_A)(1+C_B)}{(1-C_A)(1-C_B)}=e^\frac{\Phi_D}{2}.
\end{eqnarray}
while the Gauss' law in Eq. (\ref{0bc1}) takes the form of
\begin{eqnarray}
\frac{4 \epsilon_A \kappa A C_A}{C_A^2-1}-\frac{4 \epsilon_B
\kappa B C_B}{C_B^2-1}=4\pi \sigma.
\end{eqnarray}
This system of equations admits the following simple algebraic solutions for $C_A$ and $C_B$:
\begin{eqnarray}
C_A&=&\frac{\epsilon_A \kappa_A +\epsilon_B \kappa_B
\cosh{\Phi_D/2}-R}{2\pi\sigma+\epsilon_B\kappa_B
\sinh{\Phi_D/2}}. \\
C_B&=&\frac{\epsilon_B \kappa_B +\epsilon_A \kappa_A
\cosh{\Phi_D/2}-R}{-2\pi\sigma+\epsilon_A\kappa_A
\sinh{\Phi_D/2}},
\end{eqnarray}
where
\begin{eqnarray}
R^2&=&(2\pi\sigma)^2+(\epsilon_A \kappa_A)^2+(\epsilon_B
\kappa_B)^2\nonumber\\
&&+2\epsilon_A \kappa_A\epsilon_B
\kappa_B\cosh{\Phi_D/2}.\;
\end{eqnarray}
Thus, we have now determined the constants of integration explicitly in terms of the input parameters $\epsilon_A$, $\epsilon_B$, and $\Phi_D$.  As such, we can write down the
positive and negative ion densities in a straightforward manner:
\begin{eqnarray}
\rho^A_{\pm}(z)&=&\rho^A_0\left(\frac{1\pm C_Ae^{\kappa_A\, z}}{1\mp C_Ae^{\kappa_A\, z}}\right)^2,\\
\rho^B_{\pm}(z)&=&\rho^B_0\left(\frac{1\mp C_Be^{-\kappa_B\, z}}{1\pm C_B
e^{-\kappa_B\, z}}\right)^2.
\end{eqnarray}
To illustrate the profiles of the electrostatic potential $\Phi(z)$
and of the ion distributions $\rho^{\alpha}_{\pm}(z)$ across an
interface, we will now consider, in particular, the system of a
polarizable membrane between a nitrobenzene solution
[$\varepsilon^B=35.7\varepsilon_0$ at 68F] of sodium chloride and
a water solution [$\varepsilon^A=80.4\varepsilon_0$ at 68F] of
the same salt.  Given the size ratio between the ions of
$R_+/R_-\approx 0.695$, we obtain the following normalized
relations between the various Born solvation energies:
$g^A_+\approx 0.662$, $g^A_-\approx 0.446$, $ g^B_+\approx 1.44$,
and $g^B_-\equiv 1$.  Thus, it follows readily that
$\rho^A_0/\rho^B_0=e^{1.33}\approx 3.79$ and $\Phi_D \approx
0.112$, which implies that $\kappa^A/\kappa^B\approx 1.30$.  We
display the resulting plots in Figs.~\ref{potential1} and~\ref{Profiles_neg}.

\begin{figure}[t]
\begin{center}
\includegraphics{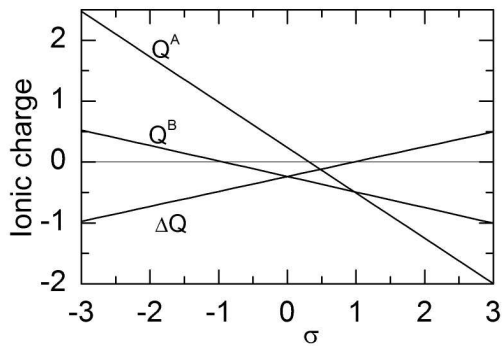}
\vspace{2mm}
\caption{Plots of the total excess charge $Q^{A}$, $Q^B$ and the net charge flux across the interface $\Delta Q$.  For this particular plot, we use a range of surface charge densities $\sigma$ in units of $\epsilon_A\,\kappa_A$ and $\Phi'_d=0.112$ for the nitrobenzene/water solution at 68F. }
\label{QT}
\end{center}
\end{figure}

The excess charge, $Q^{\alpha}_{\pm}$, defined as
\begin{eqnarray}
Q^{\alpha}_{\pm}&=&\int^{\infty}_0dz\,\left[\rho^{\alpha}_{\pm}(z)-\rho^{\alpha}_0\right]\;,
\end{eqnarray}
and due physically to the accumulation or depletion of charges near the interface as
the ions migrate across the boundary,  can also be obtained
in analytical closed-form for the present case of planar interface. We note that the excess charge $Q^{\alpha}_{\pm}$ has the dimensions of charge density per unit of transverse area.  Upon direct evaluation, we obtain
\begin{eqnarray}
Q^{A}_{\pm}&=&\frac{1}{2\pi}\frac{\epsilon_A \kappa_A C_A
}{1\mp C_A},\\
Q^{B}_{\pm}&=&-\frac{1}{2\pi}\frac{\epsilon_B \kappa_B C_B}{1\pm
C_B}.
\end{eqnarray}
The total excess charge in each phase $\alpha$ can be simply evaluated by $Q^{\alpha}=Q^{\alpha}_++Q^{\alpha}_-$. In addition, the net charge flux across the interface can be defined as
$\Delta Q=\frac12(Q^B-Q^A)$. The total net charge accumulated in
each phase compensates for the charge introduced by the polarizable
membrane to the overall system such that the relation $\sigma+Q^A+Q^B=0$ holds at all times.  The plot of $Q^A$, $Q^B$ and $\Delta Q$ for various values of the surface charge density $\sigma$ is shown in Fig. \ref{QT}.

In the case of polarizable membranes, we see that the additional surface charge density can have dramatic effects on the electrostatic potential and charge distributions of
the two phases. As shown in Fig.~\ref{potential1}, there exists a sufficiently large surface charge density on the interface, $\sigma^{*\alpha}$, at each phase $\alpha$ such that the electric field vanishes as manifested by the corresponding constant electrostatic potential. The further addition of surface
charges, either positive or negative, reverses the sign of the potential gradient and of the excess
charge in the corresponding phase.  The critical surface charge densities are found to be
\begin{eqnarray}
\sigma^{*A}&=&\frac{1}{2\pi}\epsilon_B \kappa_B
\sinh{\frac{\Phi_D}{2}},\\
\sigma^{*B}&=&-\frac{1}{2\pi}\epsilon_A \kappa_A
\sinh{\frac{\Phi_D}{2}}.
\end{eqnarray}
It can be also checked that at these critical surface charge densities,
our expressions for the excess charge $Q^{\alpha}$ and the net charge flux across the interface $\Delta Q$ also vanish in the respective phase (Fig. 4).  Notwithstanding the seeming linearity that both $Q^{\alpha}$ and $\Delta Q$ follow with respect to the surface density $\sigma$, we do observe a slight jump across $\sigma=0$ thereby establishing a small deviation to perfect linearity.

We now consider a few general properties of our solutions.
Firstly, we note that the distribution potential  $\Phi_D$  across the interface, itself generated by the
different dielectric constants of each phase, is only dependent upon ion types and not of their concentration. Thus, in the absence of any ions in the system, we must then recover a uniform potential. Indeed, we observe exactly that in approaching the limit of zero ionic concentration where the gradient of the
potential across the interface gradually decreases;  the required net jump $\Phi_D$ is being pushed outwardly towards infinity.

Another noteworthy feature is the asymptotic behavior of the potential at
large distances from the interface.  In the limit of $z\rightarrow \infty$, we have $\Phi^{A} \approx
\Phi_{D}-4 C_{A}\exp\left({\kappa_A\, z}\right)$, and $\Phi^{B} \approx 4
C_{B}\exp\left({-\kappa_B z}\right)$. Thus, we see that the integration constants now
act as potential sources in the solution form for the linearized version of the
Poisson-Boltzmann equation.  Given that each integration constant satisfies the constraint of $|C_\alpha|\leq 1$, the asymptotics always arises
from a linear source of maximum potential $\Phi_{max}\leq4$.

Lastly, we remark that our results are consistent with the Born description of solvation
energy:  since cations are in general smaller in size due to
their oxidized state, the solvation free energy is always more
negative when compared with their anionic counterpart, provided
that they are both subjected to the same solvent of the same constant
dielectric permittivity.  Therefore, it is always thermodynamically
more favorable to dissolve more of the cations in
the more polar phase and redistribute the density profiles of the
remaining ions accordingly in both phases.  The validity of our
argument remains even when the membrane becomes charged in the
same polarity as the cations.  In general, we do not expect that
the multivalency of the dissolved ions would qualitatively change
the behavior of the differential adsorption phenomenon described herein, at the
mean-field level, other than to intensify its effect as
evident in the prefactor of our Born expression of solvation
energy.

\section{Conclusions}

In this work, we have constructed a self-contained thermodynamic
formulation of ion solvation in binary immiscible polar mixtures and
described the general phenomenon of differential ion adsorption at
liquid-liquid interfaces via the full nonlinear Poisson-Boltzmann
framework. Assuming the Born model of solvation and using the
mechanics of the $\Omega$-potential, our formulation presents a
fundamental, mean-field description that relates the experimentally
detectable bulk-ion densities in each phase, as well as the
distribution potential across the membrane, to the respective
dielectric constants and sizes of each ion species present in the
two phases. We note that electrostatics alone is insufficient in
capturing the physics of ion solvation and interfacial adsorption.

Our work can be generalized in several ways.  Our formulation
models the dissolved ions specifically as point particles and the liquid-liquid
interface as sharp.  Immiscible multicomponent liquids in some cases can lead to the 
broadening of interfaces~\cite{Ching-I}.  In such instances,  adsorption of ionic component in
slabs of immiscible liquids has been recently considered using
liquid-liquid weak segregation approximations~\cite{Onuki}.  Thus, it is of
interest to consider other morphologies and systems than those presented in this work. As a refinement
to our model, it is also possible to incorporate further
molecular details in the construction of our thermodynamic
potential.  This can be done, for example, with the explicit inclusion of
short-range interactions between the ions by replacing our
expression of $\Omega^{\alpha}_{sol}$ by
\begin{eqnarray}
\Omega'^{\alpha}_{sol}&=&\int\int\,d\bm{r}d\bm{r}'G_{sr}\left(\bm{r}-\bm{r}'\right)\rho(\bm{r}')\rho(\bm{r})
\end{eqnarray}

One could then consider components with more complicated structures
than point ions, such as polyelectrolytes, and consider the
additional effects introduced by these new degrees of
freedom~\cite{Evans1979, Igal2005}. Though protein adsorption on
hard surfaces have been analyzed~\cite{Igal2003}, a general description of
macroion adsorption to liquid interfaces is lacking.  Such a general description would be highly useful, as, for example, the phenomenon of
DNA adsorption to the interface among liquid surfaces is crucial in
cell biology~\cite{Po,New}.  The inclusion of proteins and/or DNA
molecules in the system and the study of their corresponding
adsorption phenomenon along liquid-liquid interfaces will be
presented elsewhere. We point out in passing that the long-range interaction of
electrostatics would not contribute at the mean-field level due to
overall electro-neutrality.

It is our hope to have demonstrated  in this work that the
differential adsorption of ions is a general phenomenon in
electrostatics that occurs along interfaces between different
dielectric media.  Such chemical environments are ubiquitous in many
biological systems at the cellular and physiological
levels~\cite{AlbertsText}.  In particular, it was pointed out by
Onsager~\cite{Onsager} regarding the potential importance of such a
``primodial oil slick" in the early development of life.  The
phenomenon of thermodynamically driven, selective confinement of
ions and the consequent breaking of charge-conjugation symmetry near
a liquid-liquid interface would provide just the physical mechanisms
necessary for the essential organic molecules to aggregate in close,
two-dimensional proximity to each other, where the process of
diffusion would have worked much more efficiently in bringing
together these molecules and starting the chain reaction of life
than in the otherwise three-dimensional scenario in the bulk.  It is
our further hope to generalize our results in this work to other
interfacial geometries such as the cylindrical and spherical cases.

\begin{acknowledgments}

This work is supported by the ACS PRF Grant 44645-AC7 and by the
National Science Foundation grant number DMR-0414446.

\end{acknowledgments}

\end{document}